\documentclass{article}
\usepackage{cite}
\usepackage{graphicx}
\usepackage{dcolumn}
\usepackage{amsmath}

\begin{document}

\date{}
\title{On the calculation of bound-state energies supported by hyperbolic double
well potentials}
\author{Francisco M. Fern\'{a}ndez\thanks{%
fernande@quimica.unlp.edu.ar} \\
INIFTA, DQT, Sucursal 4, C.C 16, \\
1900 La Plata, Argentina}
\maketitle

\begin{abstract}
We obtain eigenvalues and eigenfunctions of the Schr\"{o}dinger equation
with a hyperbolic double-well potential. We consider exact polynomial
solutions for some particular values of the potential-strength parameter and
also numerical energies for arbitrary values of this model parameter. We
test the numerical method by means of a suitable exact asymptotic expression
for the eigenvalues and also calculate critical values of the strength
parameter that are related to the number of bound states supported by the
potential.
\end{abstract}

\section{Introduction}

\label{sec:intro}

In the last years there has been interest in quantum-mechanical models with
hyperbolic potentials\cite{HRP10,SDP12,D13,WYLYZ14,SDLDD15,HS16,KCF20}. The
reason is that some of them appear to be useful in some physical applications%
\cite{HRP10,SDP12}. Under suitable transformations the resulting eigenvalue
equations are exactly or conditionally solvable\cite
{HRP10,SDP12,D13,WYLYZ14,SDLDD15,HS16,KCF20}. The Schr\"{o}dinger equation
can be transformed into a Kummer's differential equation\cite{SDP12} or a
confluent Heun equation\cite{HRP10,D13,WYLYZ14,SDLDD15,HS16,KCF20} and the
Frobenius method (power-series approach) leads to a three-term recurrence
relation for the expansion coefficients\cite{D13,SDLDD15,HS16,KCF20}. This
fact enables one to obtain exact polynomial solutions by a suitable
truncation condition. Besides, it is also possible to obtain numerical
solutions for all the states of the problem from the same tree-term
recurrence relation through an alternative truncation condition\cite{KCF20}.

The purpose of this paper is to investigate the relationship between the
energies associated with the exact polynomial solutions and those obtained
numerically from the same three-term recurrence relation. In particular, we
are interested in the accuracy and usefulness of the numerical method. In
section~\ref{sec:model} we briefly discuss the model and the transformation
of the Schr\"{o}dinger equation into a convenient eigenvalue equation that
is suitable for the application of the Frobenius method. From a suitable
truncation condition we obtain exact polynomial solutions for some
particular bound states\cite{D13,SDLDD15,HS16,KCF20}. In section~\ref
{sec:numerical} we test a numerical method for the calculation of all the
bound-state eigenvalues from the three-term recurrence relation for the
coefficients of the Frobenius expansion\cite{KCF20}. Finally, in section~\ref
{sec:conclusions} we summarize the main results and draw conclusions.

\section{The model}

\label{sec:model}

Following Downing\cite{D13} we consider the one-dimensional Schr\"{o}dinger
equation
\begin{equation}
-\frac{\hbar ^{2}}{2m}\psi ^{\prime \prime }(x)+V(x)\psi (x)=E\psi
(x),\;V(x)=-V_{0}\frac{\sinh ^{4}\left( x/d\right) }{\cosh ^{6}\left(
x/d\right) },  \label{eq:schro}
\end{equation}
for a particle of mass $m$ in a hyperbolic potential with two model
parameters $V_{0}>0$ and $d>0$ that determine its depth and width,
respectively. This potential exhibits a barrier at $x=0$, $V(0)=0$, between
two minima at $x_{\pm }=d\,\ln \left( \sqrt{3}\pm \sqrt{2}\right) $ of depth
$V\left( x_{\pm }\right) =-4V_{0}/27$. Since $V\left( x\rightarrow \pm
\infty \right) =0$ then there will be bound-state energies in the interval $%
-4V_{0}/27<E<0$.

If we define the dimensionless coordinate $z=x/d$, the dimensionless
parameter $v_{0}=2md^{2}V_{0}/\hbar ^{2}$ and the dimensionless energy $%
\epsilon =2md^{2}E/\hbar ^{2}$ the resulting dimensionless equation\cite{F20}
\begin{equation}
-\varphi ^{\prime \prime }(z)+v(z)\varphi (z)=\epsilon \varphi
(z),\;v(z)=-v_{0}\frac{\sinh ^{4}\left( z\right) }{\cosh ^{6}\left( z\right)
},  \label{eq:schro_dim}
\end{equation}
clearly shows that there is just one relevant parameter, $v_{0}$, and not
two as some authors appeared to suggest\cite{D13,WYLYZ14,SDLDD15,KCF20}. In
fact, the problem reduces to calculating $\epsilon \left( v_{0}\right) $ and
the width parameter is only necessary in order to obtain $E\left(
V_{0},d\right) =\hbar ^{2}\epsilon \left( v_{0}\right) /\left(
2md^{2}\right) $\cite{HS16,F20}. It is clear that the dimensionless
bound-state energies will appear in the interval $-4v_{0}/27<\epsilon <0$.

According to the Hellmann-Feynman theorem\cite{G32,F39} the eigenvalues
decrease with the potential-strength parameter as
\begin{equation}
\frac{d\epsilon }{dv_{0}}=-\left\langle \frac{\sinh ^{4}\left( z\right) }{%
\cosh ^{6}\left( z\right) }\right\rangle .  \label{eq:HFT}
\end{equation}

By means of the change of variables $\xi =1/\cosh (z)^{2}$ the dimensionless
equation (\ref{eq:schro_dim}) becomes
\begin{equation}
4\xi ^{2}(1-\xi )u^{\prime \prime }(\xi )+2\xi (2-3\xi )u^{\prime }(\xi
)+\left[ \epsilon +v_{0}\xi \left( \xi -1\right) ^{2}\right] u(\xi )=0,
\label{eq:schro_xi}
\end{equation}
where $0<\xi \leq 1$. From a further transformation\cite{D13}
\begin{equation}
u(\xi )=\xi ^{\beta /2}\exp \left( \frac{\alpha }{2}\xi \right) y(\xi ),
\label{eq:u->y}
\end{equation}
where $\beta =\sqrt{-\epsilon }$ and $\alpha ^{2}=v_{0}$, we obtain the more
convenient equation
\begin{eqnarray}
&&4\xi ^{2}\left( 1-\xi \right) y^{\prime \prime }(\xi )-2\xi \left[ 2\alpha
\xi \left( \xi -1\right) +2\beta \left( \xi -1\right) +3\xi -2\right]
y^{\prime }(\xi )  \nonumber \\
&&-\left[ \alpha ^{2}\left( \xi -1\right) +\alpha \left( 2\beta \left( \xi
-1\right) +3\xi -2\right) +\beta ^{2}+\beta \right] y(\xi )=0.
\label{eq:y(xi)}
\end{eqnarray}
It follows from the bounds to the dimensionless energies discussed above
that
\begin{equation}
0<\beta <\frac{2|\alpha |}{\sqrt{27}}.  \label{eq:beta_bounds}
\end{equation}

The solution $y(\xi )$ can be expanded in a Taylor series about the origin
\begin{equation}
y(\xi )=\sum_{j=0}^{\infty }c_{j}\xi ^{j},  \label{eq:y_series}
\end{equation}
and the expansion coefficients $c_{j}$ satisfy the three-term recurrence
relation
\begin{eqnarray}
c_{j+2}(\alpha ,\beta ) &=&A_{j}(\alpha ,\beta )c_{j+1}(\alpha ,\beta
)+B_{j}(\alpha ,\beta )c_{j}(\alpha ,\beta ),  \nonumber \\
j &=&-1,0,1,2,\ldots ,\;c_{-1}=0,\;c_{0}=1,  \nonumber \\
A_{j}(\alpha ,\beta ) &=&-\frac{\left( \beta +2j+3\right) \left( 2\alpha
-\beta -2\left( j+1\right) \right) +\alpha ^{2}}{4\left( \beta +j+2\right)
\left( j+2\right) },  \nonumber \\
B_{j}(\alpha ,\beta ) &=&\frac{\alpha \left( \alpha +2\beta +4j+3\right) }{%
4\left( \beta +j+2\right) \left( j+2\right) }.  \label{eq:TTRR_1}
\end{eqnarray}
For those physically acceptable solutions to the truncation conditions $%
c_{n}\neq 0$, $c_{n+1}=0$ and $c_{n+2}=0$, $n=0,1,\ldots $, the series (\ref
{eq:y_series}) reduces to a polynomial of degree $n$. It follows from these
conditions that $B_{n}(\alpha ,\beta )=0$ which forces a relationship
between $\alpha $ and $\beta $. We arbitrarily choose
\begin{equation}
\beta =\beta _{n}=-\frac{\alpha +4n+3}{2},  \label{eq:beta(alpha)}
\end{equation}
so that
\begin{eqnarray}
A_{j,n}(\alpha ) &=&-\frac{\alpha ^{2}-8\alpha \left( 3j-3n+2\right) +\left(
4j-4n+1\right) \left( 4j-4n+3\right) }{8\left( \alpha -2j+4n-1\right) \left(
j+2\right) },  \nonumber \\
B_{j,n}(\alpha ) &=&\frac{2\alpha \left( n-j\right) }{\left( \alpha
-2j+4n-1\right) \left( j+2\right) }.  \label{eq:Aj_Bj_n}
\end{eqnarray}

The coefficient $c_{n+1}$ is a rational function of $\alpha $ and its
numerator is a polynomial of degree $2(n+1)$; therefore, the remaining
condition $c_{n+1}=0$ has $2(n+1)$ solutions $\alpha _{n,i}$, $i=1,2,\ldots
,2(n+1)$. Numerical calculation suggests that all the roots are real;
however, not all of them are physically acceptable. It follows from
equations (\ref{eq:beta_bounds}) and (\ref{eq:beta(alpha)}) that there are
exact polynomial solutions only for those roots that satisfy
\begin{equation}
-\frac{27+12\sqrt{3}}{11}\left( 4n+3\right) <\alpha _{n,i}<-\left(
4n+3\right) .  \label{eq:alpha_n,i_bounds}
\end{equation}
The polynomial solutions to equation (\ref{eq:y(xi)}) are of the form
\begin{equation}
y^{(n,i)}(\xi )=\sum_{j=0}^{n}c_{j}\left( \alpha _{n,i}\right) \xi ^{j},
\label{eq:y_poly}
\end{equation}
for those values of $\alpha _{n,i}$ in the interval given in equation (\ref
{eq:alpha_n,i_bounds}). It is worth noticing that present hyperbolic
potential supports at least one bound state for any positive value of $v_{0}$
( see, for example, \cite{HS16} and references therein).

Since $\xi $ is an even function of $z$ then all the solutions obtained in
the way just described are even functions of $z$. It is obvious that there
should be even $\varphi ^{e}(z)$ and odd $\varphi ^{o}(z)$ solutions to the
dimensionless eigenvalue equation (\ref{eq:schro_dim}) and the approach just
outlined is unable to provide the latter. An alternative strategy is based
on the more convenient variable $-1<\zeta =\tanh (z)<1$ that enables us to
obtain both even and odd solutions\cite{D13}. However, it is possible to
obtain also the odd solutions from the transformation just discussed. In
fact, since $\varphi ^{o}(0)=0$ and $\xi (0)=1$ it is only necessary to
force a zero at $\xi =1$ as discussed by Wen et al\cite{WYLYZ14} and Hall
and Saad\cite{HS16}. This more general approach is outlined in Appendix~\ref
{sec:even_odd}.

Figure~\ref{Fig:HDW1} shows that the highest roots $\epsilon ^{(n,i)}$ (for
the exact polynomial solutions) follow a neat decreasing curve in terms of $%
\alpha _{n,i}^{2}$. However, this curve has no physical meaning as shown in
what follows.

\section{Numerical calculation}

\label{sec:numerical}

The truncation condition discussed in the preceding section only yields some
particular energies $\epsilon ^{(n,i)}$ for some particular values of the
strength parameter $v_{0}^{(n,i)}=\alpha _{n,i}^{2}$, provided that the
values of $\alpha _{n,i}$ satisfy the bounds in equation (\ref
{eq:alpha_n,i_bounds}). Such results are almost useless if one is not able
to identify and organize them properly. Kufel et al\cite{KCF20} proposed an
approach for the calculation of all the bound-state energies that is similar
to a procedure developed some time ago by Myrheim et al\cite{MHV92}. The
strategy consists of setting the desired value of $\alpha $, calculating the
coefficients $c_{j}$, $j=0,1,\ldots ,N$, from the recurrence relation (\ref
{eq:TTRR_1}) and then solving the equation $c_{N}=0$ for $\beta $. Those
sequences of roots $\beta ^{(N,j)}$, $N=N_{I},N_{I}+1,\ldots $ that converge
to a limit in the range given by equation (\ref{eq:beta_bounds}) are
expected to yield the energy eigenvalues for the chosen value of $\alpha $.

\begin{figure}[tbp]
\begin{center}
\includegraphics[width=9cm]{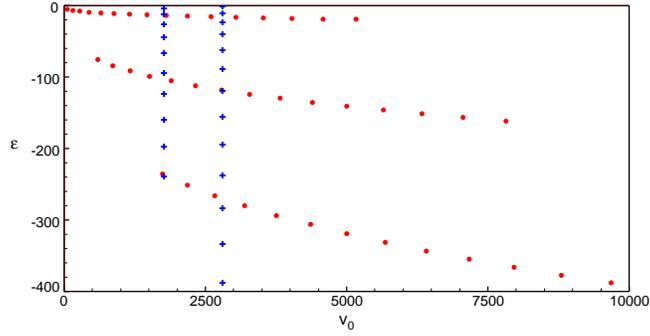}
\end{center}
\caption{Eigenvalues from the truncation method (red points) and numerical
ones for $v_0=1764$ and $v_0=2809$ (blue crosses)}
\label{Fig:HDW2}
\end{figure}

Figure~\ref{Fig:HDW2} shows numerical eigenvalues for two values of $v_{0}$
(blue crosses). This figure also shows the exact eigenvalues stemming from
the truncation condition discussed in the preceding section (red circles).
Notice that the exact eigenvalues follow well defined curves (the highest
one shown in Figure~\ref{Fig:HDW1} in a smaller scale) which, in principle,
do not have any physical meaning because they connect bound states with
different quantum numbers.

In order to test the validity of this numerical calculation of the energies
supported by the hyperbolic double well we resort to a simple asymptotic
expression, valid for sufficiently large values of $v_{0}$. Under such
condition we expand the potential in a Taylor series about either minima
\begin{equation}
V(z)=v_{0}\left[ -\frac{4}{27}+\frac{8}{27}\left( z-z_{\pm }\right)
^{2}+\ldots \right] ,\;z_{\pm }=\ln \left( \sqrt{3}\pm \sqrt{2}\right) ,
\label{eq:V_Taylor}
\end{equation}
and apply the harmonic approximation. In this way we obtain approximate
asymptotic eigenvalues
\begin{equation}
\epsilon _{\nu }^{asymp}=-\frac{4}{27}v_{0}+2\sqrt{\frac{2v_{0}}{27}}\left(
2\nu +1\right) ,\;\nu =0,1,\ldots ,k.  \label{eq:epsilon_asymp}
\end{equation}

\begin{figure}[tbp]
\begin{center}
\includegraphics[width=9cm]{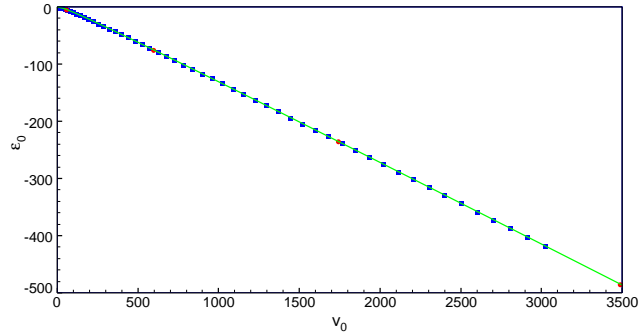}
\end{center}
\caption{Ground state calculated numerically (blue squares), by means of the
exact truncation condition (red points) and from the asymptotic expression (%
\ref{eq:epsilon_asymp}) (green line)}
\label{Fig:e0}
\end{figure}

Figure~\ref{Fig:e0} shows that the numerical values of $\epsilon _{0}$ (blue
squares) already agree with $\epsilon _{0}^{asymp}$ (green line) which
strongly suggests that the truncation condition $c_{N}=0$ is suitable for
obtaining the eigenvalues of the hyperbolic double well ($N=10$ was
sufficient for the scale of this figure). This figure also shows some exact
results (red circles) given by the truncation condition of the preceding
section. As a further test of the method just outlined we have also verified
that the energies obtained in this way agree with those coming from the
widely tested Riccati-Pad\'{e} method\cite{FMT89a}.

According to the Hellmann-Feynman theorem (\ref{eq:HFT}) the bound-state
energies $\epsilon _{\nu }$ decrease with $v_{0}$. For any value of $v_{0}$
there is at least one bound state of even symmetry and the number of bound
states supported by the potential increases with $v_{0}$ (see, for example,
Hall and Saad\cite{HS16} and references therein for a discussion of this
issue). Consequently, there are critical values of the potential-strength
parameter $v_{0}=v_{0,K}$, $K=1,2,\ldots $, such that $\epsilon _{K}=0$.
Their meaning is that there is just one bound state for $v_{0}<v_{0,1}$ and $%
K+1$ bound states for $v_{0,K}<v_{0}<v_{0,K+1}$. The numerical method
outlined above enables us to obtain the critical values $\alpha _{K}$ of $%
\alpha $ in a simple way. We simply set $\beta =0$ and solve $c_{N}=0$ for $%
\alpha $ for sufficiently large values of $N$. From a straightforward
calculation with $N\leq 36$ we estimated $\alpha _{2}=-5.272715881$, $\alpha
_{4}=-9.398121349$, $\alpha _{6}=-13.455570$ and $\alpha _{8}=-17.4897$. On
using the three-term recurrence relation with the factors given in equation (%
\ref{eq:TTRR_1_eo}) and $\gamma =1$ we obtain $\alpha _{1}=-2.073164811$, $%
\alpha _{3}=-6.181847266$, $\alpha _{5}=-10.22002699$, $\alpha
_{7}=-14.2405704$ and $\alpha _{9}=-18.25373$.

\section{Conclusions}

\label{sec:conclusions}

In this paper we have discussed the exact polynomial solutions to the
Schr\"{o}dinger equation with the hyperbolic double well potential shown in
equation (\ref{eq:schro_dim}). In particular, we focused on the bounds to
the physically acceptable values $\alpha_{n,i}$ of the parameter $\alpha =-%
\sqrt{v_{0}}$ and showed that the energies $\epsilon ^{(n,i)}$ appear on
some decreasing curves on the $v_{0}-\epsilon $ plane, though their meaning
is not clear to us. We also compared these exact eigenvalues with some
numerical results $\epsilon _{\nu }\left( v_{0}\right) $ provided by a
simple method based on the three-term recurrence relation\cite{KCF20}. In
order to test the validity of the numerical approach we resorted to an exact
asymptotic expression for the eigenvalues, valid for sufficiently deep
wells. The numerical approach proves useful for the calculation of the
critical values $v_{0,K}=\alpha _{K}^{2}$ of the strength parameter $v_{0}$
that are related to the number of bound states supported by the potential.

\appendix

\numberwithin{equation}{section}

\section{Even and odd states}

\label{sec:even_odd}

In order to obtain odd states from the procedure outlined in section~\ref
{sec:model} we have to force a zero at $\xi =1$, which can be easily done by
the transformation\cite{WYLYZ14,HS16}
\begin{equation}
u(\xi )=\xi ^{\beta /2}\left( 1-\xi \right) ^{\gamma /2}\exp \left( \frac{%
\alpha }{2}\xi \right) y(\xi ),  \label{eq:u->y_eo}
\end{equation}
where $\beta =\sqrt{-\epsilon }$, $\alpha ^{2}=v_{0}$ (as before) and $%
\gamma (\gamma -1)=0$. When $\gamma =0$ we recover the ansatz of section~\ref
{sec:model} for even states and $\gamma =1$ gives us the odd states. The
differential equation becomes
\begin{eqnarray}
&&4\xi ^{2}\left( 1-\xi \right) y^{\prime \prime }(\xi )-2\xi \left[ 2\alpha
\xi \left( \xi -1\right) +2\beta \left( \xi -1\right) +2\gamma \xi +3\xi
-2\right] y^{\prime }(\xi )  \nonumber \\
&&-\xi \left\{ \alpha ^{2}\left( \xi -1\right) +\alpha \left[ 2\beta \left(
\xi -1\right) +2\gamma \xi +3\xi -2\right] +\beta ^{2}+\beta \left( 2\gamma
+1\right) \right.   \nonumber \\
&&\left. +\gamma (\gamma +1)\right\} y(\xi ).  \label{eq:y(xi)_eo}
\end{eqnarray}
On arguing as in section~\ref{sec:model} we obtain a similar three-term
recurrence relation with
\begin{eqnarray}
A_{j}(\gamma ,\alpha ,\beta ) &=&-\frac{\alpha ^{2}+2\alpha \left( \beta
+2j+3\right) -\beta ^{2}-\beta \left( 2\gamma +4j+5\right) }{4\left( \beta
+j+2\right) \left( j+2\right) }  \nonumber \\
&&+\frac{\gamma ^{2}+\gamma \left( 4j+5\right) +2\left( j+1\right) \left(
2j+3\right) }{4\left( \beta +j+2\right) \left( j+2\right) },  \nonumber \\
B_{j}(\gamma ,\alpha ,\beta ) &=&\frac{\alpha \left( \alpha +2\beta +2\gamma
+4j+3\right) }{4\left( \beta +j+2\right) \left( j+2\right) }.
\label{eq:TTRR_1_eo}
\end{eqnarray}
Upon choosing
\begin{equation}
\beta =\beta _{n}=-\frac{\alpha +2\gamma +4n+3}{2},
\label{eq:beta(alpha)_eo}
\end{equation}
we have
\begin{eqnarray}
A_{j,n}(\gamma ,\alpha ) &=&-\frac{\alpha ^{2}+8\alpha \left( \gamma
-3j+3n-2\right) +(4j-4n-1)(4j-4n+3)}{8\left( \alpha +2\gamma -2j+4n-1\right)
\left( j+2\right) },  \nonumber \\
B_{j,n}(\gamma ,\alpha ) &=&\frac{2\alpha \left( n-j\right) }{\left( \alpha
+2\gamma -2j+4n-1\right) \left( j+2\right) },  \label{eq:Aj_Bj_n_eo}
\end{eqnarray}
and the tree-term recurrence relation yields exact polynomial solutions for $%
y(\xi )$. Only the roots $\alpha _{n,i}$ of $c_{n+1}=0$ that satisfy
\begin{equation}
-\frac{27+12\sqrt{3}}{11}\left( 4n+3+2\gamma \right) <\alpha _{n,i}<-\left(
4n+3+2\gamma \right) ,  \label{eq:alpha_n,i_bounds_eo}
\end{equation}
are physically acceptable. These bounds apply only to the exact polynomial
solutions.

We can also obtain numerical eigenvalues for both even and odd states and
any $v_{0}>0$ from the roots of $c_{N}=0$ as indicated in section~\ref
{sec:numerical}.

\begin{figure}[tbp]
\begin{center}
\includegraphics[width=9cm]{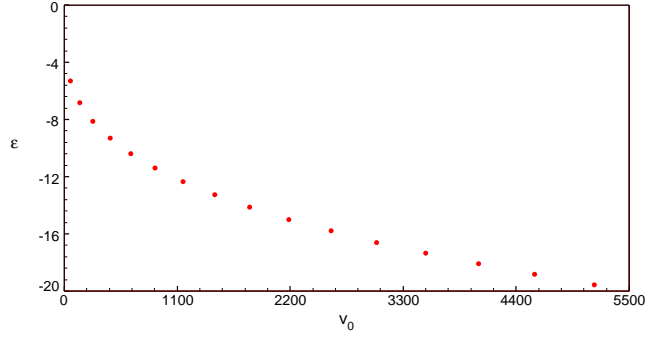}
\end{center}
\caption{Highest eigenvalues for polynomial solutions (\ref{eq:y_poly}) }
\label{Fig:HDW1}
\end{figure}

\end{document}